\documentclass[12pt,draftcls,onecolumn]{IEEEtran}

\ifCLASSINFOpdf
\else
\fi

\usepackage{amsmath}
\usepackage{bbm}
\usepackage{amssymb}
\usepackage{graphicx}
\usepackage{algorithm}
\usepackage{algorithmicx}
\usepackage{algpseudocode}
\usepackage[noadjust]{cite}
\newtheorem{theorem}{Theorem}

\begin{document}

\title{Optimal Bidding in Repeated Wireless Spectrum Auctions with Budget Constraints}

\author{\IEEEauthorblockN{Mehrdad Khaledi and Alhussein A. Abouzeid}\\
\IEEEauthorblockA{Department of Electrical, Computer and Systems Engineering\\
Rensselaer Polytechnic Institute\\
Troy, NY 12180-3590, USA\\
Email: khalem@rpi.edu, abouzeid@ecse.rpi.edu}
}
\maketitle

\begin{abstract}
Small operators who take part in secondary wireless spectrum markets typically have strict budget limits. In this paper, we study the bidding problem of a budget constrained operator in repeated secondary spectrum auctions. In existing truthful auctions, truthful bidding is the optimal strategy of a bidder. However, budget limits impact bidding behaviors and make bidding decisions complicated, since bidders may behave differently to avoid running out of money. We formulate the problem as a dynamic auction game between operators, where knowledge of other operators is limited due to the distributed nature of wireless networks/markets. We first present a Markov Decision Process (MDP) formulation of the problem and characterize the optimal bidding strategy of an operator, provided that opponents' bids are i.i.d. Next, we generalize the formulation to a Markov game that, in conjunction with model-free reinforcement learning approaches, enables an operator to make inferences about its opponents based on local observations. Finally, we present a fully distributed learning-based bidding algorithm which relies only on local information. Our numerical results show that our proposed learning-based bidding results in a better utility than truthful bidding.
\end{abstract}


\section{Introduction}
\IEEEPARstart{S}{pectrum} scarcity has become a major challenge due to the rapid growth in mobile wireless communications. Several measurement studies indicate that the problem lies in inefficient use of the available wireless spectrum rather than scarcity of the spectrum \cite{FCCRep}. Secondary spectrum markets have emerged to improve the spectrum utilization, where a primary license owner (PO) can lease its idle spectrum band(s) to unlicensed secondary users (SU) for a short period of time. A common approach for leasing the spectrum is holding an auction among SUs.

Several auction mechanisms have been proposed in the literature for re-allocating the spectrum in secondary markets that mostly focus on a single round of auction (one-shot mechanisms) \cite{Khaledi13,Hoefer15}. However, secondary spectrum auctions repeat frequently, since spectrum access is granted for a short period of time. The difficulty arises as the SUs can learn some information about their opponents and the environment over time, which consequently complicate their bidding decisions.

In a repeated auction environment, the major problem of an SU is to find an optimal bidding strategy that maximizes its long-term utility. The decision making process of SUs in a repeated spectrum auction is studied in \cite{Han11}. In their model, SUs choose between participating in the auction by bidding their true valuations, or staying out of the auction to just monitor the results. Assuming independent and identically distributed (i.i.d.) SUs' valuations, a threshold is derived for SUs above which they should participate in the auction. In a more general context, \cite{Akkarajitsakul11} utilizes Bayesian auction games for resource allocation in wireless networks, which entails maintaining beliefs about private information of others. Similarly, \cite{Ji08} presents a multi-stage double auction game, however, it only solves for a single round of auction. \cite{Bae08} presents a sequential bandwidth and power auction among SUs. A unique equilibrium is guaranteed for the case of two SUs, provided that full information about private valuations are available. From a PO perspective, the spectrum pricing competition has been studied extensively \cite{Duan10,Niyato08,Niyato09,Min12}. In such models, SUs choose a spectrum provider solely based on the offered price, then bid their true valuations.

We study repeated spectrum auctions in presence of budget constrained SUs, since in real world scenarios, bidders have limits on the amount of money they can spend. According to an analysis of FCC's spectrum auctions \cite{Cramton97}, many local wireless operators have budget limits, and these limits affect their bidding behaviors. Each operator typically starts with an initial budget to invest in the spectrum market. The operator improves its utility after winning an auction and getting high quality channel access for its services. In case of losing an auction, the operator may resort to opportunistic generalized access mode which does not provide a quality of service guarantee, \cite{FCCRep35GHz}. Therefore, an operator needs to bid wisely and plan its budget to get the most value from its participation in multiple rounds of auction. It should be noted that we use the terms operator and SU interchangeably in this paper.

Our goal is to characterize an optimal bidding strategy for a budget constrained SU in repeated secondary spectrum auctions. To the best of our knowledge, optimizing the bidding strategy of an SU in presence of budget limits has not been previously considered in the literature. The challenge presented by budget limits is that it makes the bidding decisions complex, since an SU needs to take into account both the competition in the market and its own budget constraints. \emph{Truthful bidding is no longer the optimal bidding strategy when SUs are budget constrained, as SUs may behave differently to avoid running out of money.} Thus, in contrast with prior works \cite{Han11,Fu09} that assume SUs always bid their true valuations, budget constrained SUs have a wide variety of strategies for bidding. Therefore, SUs face a budget planning problem and they need to find utility-maximizing bids without exceeding their budgets.

The significance of our work is that we propose solutions for the bidding problem of a budget constrained SU, with and without i.i.d competing bids. We characterize the optimal bidding strategy of an SU, when opponents' bids are i.i.d. For the case when no information about other SUs is available, we present a learning-based bidding algorithm that relies only on local information, and is well-suited to wireless environments/markets. It is worth noting that budget optimization has been studied in the context of online keyword advertising. For instance, \cite{Borgs07} and \cite{Feldman07} analyze random bids and present bidding heuristics for advertisers to maximize their return on investments. Also, \cite{Amin12} proposes a greedy algorithm for budget optimization with a single keyword and a single advertising slot. Similarly, \cite{Gummadi13} studies the bidding problem for a single keyword assuming a bidder faces large (theoretically infinite) number of i.i.d bidders. However, such an assumption does not typically hold in the context of wireless spectrum markets, since there are limited number of competing SUs.

It should also be noted that our approach is different from the literature of dynamic auction design, where the objective is to design efficient or revenue-optimal mechanisms in dynamic environments (e.g. \cite{Khaledi15}). Instead of designing a complex mechanism that focuses on the PO's side, we consider repetition of simple auction mechanisms, and we study the dynamics of such a system from SUs' point of view. In this setting, we analyze the bidding strategies of an SU. In fact, an SU is faced with a trade-off between the possibility of getting a surplus in the current auction and the possibility of getting a larger but uncertain surplus in future auctions, subject to its budget limit.

In this paper, we make the following contributions. We formulate the budget-constrained spectrum sharing problem as a repeated auction game in which SUs compete to get one of the available channels. We first present a Markov Decision Process (MDP) formulation of the problem and characterize the optimal bidding strategy of an SU, assuming that opponents' bids are i.i.d. Next, we generalize the formulation to a Markov game, where an SU can make inferences about its competitors based on its local observations, and i.i.d. bids assumption is not required. Finally, we present a fully distributed learning-based bidding algorithm which relies only on local information.

The rest of this paper is organized as follows. Section~\ref{systemmodel} presents the system model used in this paper. In Section~\ref{SecFormulation}, we present a formulation of the optimal bidding problem of a budget constrained SU. We characterize the optimal bidding strategy of an SU, assuming that the SU faces i.i.d opponent bids in Section~\ref{Seciid}. In Section~\ref{SecLearning}, we present a fully distributed learning-based bidding algorithm for an SU, which does not require i.i.d. bids assumption. Numerical results are presented in Section~\ref{simulation}. Finally, Section~\ref{conclusion} concludes the paper and outlines possible avenues for future work.

\section{System Model}\label{systemmodel}
We consider a network consisting of a set of secondary users/operators (SUs) who are willing to buy channel access for their services from a primary spectrum owner (PO). SUs are budget constrained and compete with other SUs in a repeated auction where the PO acts as the auctioneer, and SUs are the bidders. The auction is repeated over time which is indexed by $t=0, 1, 2, \cdots$. We assume that each SU can get at most one of the $k$ available channels, and that each channel can be leased to one SU at each time slot.

An SU's valuation for a channel is the benefit for that specific SU of obtaining that channel. Similar to \cite{Bae08,Min12,Khaledi13}, the SUs' valuations for a channel can be related to the achievable capacity of that channel. Let $W$ be the channel bandwidth, $P_0$ be the transmission power, $N_0$ be the power spectral density of the additive noise, and let $G_i$ denote the channel gain for SU $i$. The valuation of SU $i$ for channel access, $v_i$, can be defined as:
\begin{equation}\label{equValuation}
v_i \triangleq \theta_i \; W \log(1+\frac{P_0\;G_i}{N_0\;W}) ,
\end{equation}
\noindent where $\theta_i$ is a real number which reflects the urgency of channel access for SU $i$, the more urgent the channel access to SU $i$; the higher the monetary value $\theta_i$. SUs can set their $\theta$s based upon their service types. For delay sensitive multimedia applications they have a different urgency than delay tolerant services.

It is worth noting that the model presented in this paper works with other valuation functions, and (\ref{equValuation}) is one example of such a function. We assume that at each time step, each SU can observe its current valuation, and that valuations evolve according to a Markov probability model. Let $v_i^t$ denote the valuation of SU $i$ at time $t$, then $P(v_i^{t+1}|v_i^t, v_i^{t-1},\cdots , v_i^0)=P(v_i^{t+1}|v_i^t)$. Each SU knows its own valuation probability transition model which can be learnt over time. \cite{Khaledi15} presents a model in which SUs learn their valuations over time.

In this paper, we utilize the well-known Vickrey-Clarke-Groves (VCG) auction \cite{AGT} in each round. At time step $t$, the VCG mechanism takes the SUs' bids as input and determines the output for each SU as
\begin{equation}\label{auctionOutput}
    o_i^t=\{(x_i^t,p_i^t) | x_i^t\in \{0,1\} \wedge \sum_i x_i^t\leq k\}, \forall i,
\end{equation}

\noindent where the output consists of the allocation indicator, which determines whether a channel is allocated to SU $i$ or not, and the payment that SU $i$ needs to make.

According to the VCG mechanism, $k$ identical channels are allocated to the SUs with $k$ highest bids. The winning SUs need to pay the externality\footnote{In other words, an SU pays the difference between the social welfare of the others with and without its participation \cite{AGT}.} that they cause on other SUs. Since channels are identical, the winners pay the $(k+1)$th highest bid. Therefore, we have $p_i^t=p^t=(k+1)$th highest bid if $x_i^t=1$, and $p_i^t=0$ otherwise. In such an auction, $(k+1)$th highest bid is a \emph{threshold} bid for winning the auction and winners pay that threshold.

The auction mechanism in each step can be summarized as follows. First, SU $i$ observes its valuation $v_i^t$. Second, SU $i$ decides what to bid in the current round which is denoted by $b_i^t$. Third, The PO holds the auction based on the VCG mechanism. Finally, SU $i$ observes its bidding result $o_i^t$, defined in (\ref{auctionOutput}).

We focus on the bidding problem faced by an SU in the described repeated auction environment. At each time step, an SU's bid depends not only on its valuation, but also on its remaining budget and the behavior of its competitors. In conventional auction settings, where SUs are not budget constrained, it is in SUs' best interest to bid their true valuations. Thus, truthful bidding is the best strategy of an SU regardless of its opponents. However, in presence of budget limits, truthful bidding is no longer the best strategy. For instance, consider an SU with valuation of $6$ and budget of $6$ at time $t$, when the winning threshold is $5$. Following truthful bidding, the SU bids $6$, wins the channel and gets a utility of $1$. Assuming the SU makes a fixed income of $1$, its remaining budget for time $t+1$ equals $2$. Suppose at time $t+1$, the SU's valuation and the winning threshold are $7$ and $3$, respectively. Obviously, the SU does not have enough budget to win in this round. However, the SU could have underbid at time $t$ to save its budget for time $t+1$, where it could get a utility of $4$. In fact, this simple example shows that an SU needs to plan its budget and find its optimal bidding strategy accordingly. In addition, the SU needs to take into account the behavior of its opponents in its decision making process. However, due to the distributed nature of network, knowledge about other SUs is limited, and each SU may learn some information about its opponents by repeatedly participating in the auction.

\begin{figure}[!t]
\centering
\includegraphics[width=10cm,height=7.8cm]{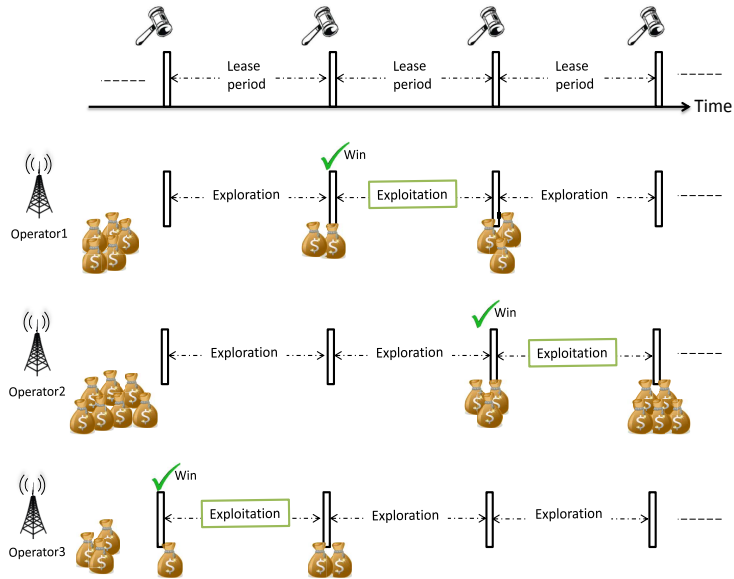}
\caption{Three budget constrained operators compete for a single channel in a repeated auction. If an operator wins, it can exploit the channel and earn an income. By participating in several rounds of auction, operators get the chance to explore and learn about their opponent's bids.}
\label{auctionGame}
\end{figure}

An instance of the problem setting is depicted in Figure~\ref{auctionGame} where three secondary operators compete for channel access in a repeated auction. Each SU typically starts with an initial budget to invest in the spectrum market. An SU improves its utility after winning an auction and getting channel access for its services. At each time step, an SU can explore and learn more information about its opponents or exploit the current information and bid to win a channel.

\section{Optimal Bidding Problem Formulation}\label{SecFormulation}

In this section, we formulate the optimal bidding problem of a budget constrained SU in the repeated auction environment described in Section~\ref{systemmodel}. Let $m_i^t$ be the remaining budget of SU $i$ at time $t$. SU $i$ observes its valuation at time $t$ and places a bid $b_i^t$, which results in an immediate utility of $(v_i^t - p_t) \mathbbm{1}_{p_t < b_i^t \leq m_i^t}$. The SU's problem is to find a bidding strategy that maximizes its long-term discounted utility. The SU's objective at time $t$ can be written as:

\begin{equation}\label{EQ-objective}
    \mathbb{E}\big[\sum_{t'=t}^{\infty} \delta^{t'-t} \; (v_i^{t'} - p_{t'}) \mathbbm{1}_{p_{t'} < b_i^{t'} \leq m_i^{t'}}\big]
\end{equation}

\noindent where the expectation is taken over the winning threshold, $p_{t'}$, and $0\leq \delta <1$ is the discount factor that controls how important future rewards are in current decisions (with larger values of $\delta$ giving more weight to future situations, as opposed to immediate rewards).

The bidding problem of SU $i$ can be modeled as a Markov decision process (MDP) that is described by a quadruple \mbox{$<S_i, B_i, q_i, r_i>$}. $S_i$ corresponds to a finite set of states of SU $i$, where state of an SU is specified by its valuation and its remaining budget. Formally, the state of SU $i$ at time $t$ is defined as $s_i^{t}=(m_i^{t}, v_i^{t})$. $B_i$ denotes a finite set of actions for SU $i$, where an action corresponds to placing a bid, $b_i^{t}$. State transition probability for SU $i$ is represented by $q_i$. Therefore, $q_i(s_i^{t+1}| s_i^{t}, o_i^t)$ is the probability that the state of SU $i$ changes from $s_i^t$ to $s_i^{t+1}$ when the auction output is $o_i^t$. State of an SU transitions as follows. SU $i$'s valuation is drawn i.i.d over time, and its budget evolves according to the following equation

\begin{equation}\label{equStateTransition}
    m_i^{t+1}=
    \begin{cases}  m_i^{t} + a_i - p_t, & x_i^t=1 \\
    m_i^{t},  & \text{Otherwise}
    \end{cases}
\end{equation}

\noindent where $a_i$ denotes a fixed income that the SU earns from getting channel access permission, and $p_t$ is the threshold amount that the SU pays for winning the auction at time $t$.

The immediate reward of SU $i$ in an auction round is denoted by $r_i$, which is the difference between SU's valuation and its payment if the SU wins a channel. 

\begin{equation}\label{equImReward}
    r_i^t (s_i^{t}, b_i^{t}, o_i^t)=
    \begin{cases}  v_i^t - p_t, & x_i^t=1 \\
    0,  & \text{Otherwise}
    \end{cases}
\end{equation}

With the MDP formulation, the SU $i$'s objective is to find a stationary strategy $\pi_i$ that maps its current state (valuation and remaining budget) into a bid to maximize its long-term discounted utility given by

\begin{equation}\label{EQ-objectiveSimple}
    \max_{\pi_i \in \Pi_i}\:\mathbb{E}\big[\sum_{t'=t}^{\infty} \delta^{t'-t} \;  r_i^{t'} (s_i^{t'}, b_i^{t'}, o_i^{t'})].
\end{equation}

\section{Optimal Bidding with i.i.d SUs}\label{Seciid}

In this section, we characterize the optimal bidding strategy of an SU assuming that bids of SUs are independent and identically distributed (i.i.d.). This assumption implies that the SU knows the probability distribution of the winning threshold. While the assumption of i.i.d bidders is common in the prior work \cite{Han11}, in Section~\ref{SecLearning}, we present a learning-based approach that does not require i.i.d opponent bids.

We define the value function of the described MDP as the maximum (over all bidding strategies) expected discounted utility of an SU. Let $U(m)$ be the value function starting with budget $m$, using the dynamic programming principle we can write

\begin{multline}\label{EQvaluefunc}
    U(m)=\mathbb{E}_v \bigg[ \max_{b \leq m}\:\mathbb{E}_p\big[ (v - p + \delta U(m-p+a) ) \mathbbm{1}_{p < b} + \delta U(m) \mathbbm{1}_{p \geq b} \big]\bigg]
\end{multline}

The SU wins if it bids strictly more than the winning threshold $p$. In this case, the SU gets an immediate reward of $v-p$ in addition to the discounted expected future utility of starting with budget $m-p+a$. If the SU loses the auction, it gets the discounted expected utility with the same initial budget. It is worth noting that since we consider the bidding problem of a typical SU, we omit the SU index for simplicity of notation. Also, we leave out the time index in the above recursive formula.

For every possible winning threshold $p$, the SU's optimal bid can be found by simulating a single-shot VCG auction in which the winning threshold is represented by a function $f$ defined as:

\begin{equation}\label{EQfunctiondef}
    f(p,m)= p + \delta (U(m) - U(m-p+a)).
\end{equation}

\noindent The function $f$ defines the costs associated with winning a round of auction. The first term $p$ is the immediate cost that the winning SU needs to pay. The second term in (\ref{EQfunctiondef}) is the exploitation cost which is incurred when the SU wins the current round of auction and starts the next round with budget $m-p+a$, compared to the case of losing the current auction and starting the next round with the same budget. In fact, exploitation cost is the discounted utility difference between winning and not winning the current round of auction.

Now, the optimal bid can be defined as a function of the current state (consisting of budget and valuation) as follows:

\begin{equation}\label{EQoptimal bid}
    b^*(m,v)=\arg\max_{b \leq m} \mathbb{E}_p \big[ (v - f(p,m)) \mathbbm{1}_{p < b \leq m}\big].
\end{equation}

\noindent In the following theorem, we characterize the optimal bid of an SU.

\begin{theorem}\label{theorem1}
The optimal bidding strategy of a budget constrained SU in the described repeated VCG auction (Section~\ref{systemmodel}) is characterized as
\begin{equation*}
    b^*(m,v)=\min (m, f^{-1}(v,m))
\end{equation*}

\noindent where $f^{-1}(v,m)$ is the $z$ such that $f(z,m)=v$.
\end{theorem}
\begin{IEEEproof}[Proof]
The main idea is to transform the current auction round into a single-shot VCG auction where the winning threshold is represented by the function $f(p,m)$ (\ref{EQfunctiondef}). It should be noted that by definition, $f(p,m)$ is strictly increasing in $p$. Therefore, the following two conditions are equal,
\begin{equation*}
    \mathbbm{1}_{p < b \leq m} = \mathbbm{1}_{f(p,m) < f(b,m) \leq f(m,m)}.
\end{equation*}
\noindent Now we can rewrite the optimal bid function (\ref{EQoptimal bid}) as
\begin{equation}\label{EQProofoptimal bid}
    b^*(m,v)=\arg\max_{b} \mathbb{E}_p \big[ (v - f(p,m)) \mathbbm{1}_{f(p,m) < f(b,m) \leq f(m,m)}\big]
\end{equation}

The optimal bid of an SU in a single-shot VCG auction is the SU's valuation subject to its budget limit which can be represented by $\min(v,m)$. Therefore, for a single-shot VCG, we can write
\begin{equation}\label{EQProofSingleVCG}
    \arg\max_{b} \mathbb{E}_p \big[ (v - p) \mathbbm{1}_{p < b \leq m}\big]=\min(v,m).
\end{equation}

\noindent We can replace $m$ by $f(m,m)$ and $p$ by $f(p,m)$ in (\ref{EQProofSingleVCG}),
\begin{equation*}
    \arg\max_{b} \mathbb{E}_p \big[ (v - f(p,m)) \mathbbm{1}_{f(p,m) < b \leq f(m,m)}\big] = \min(v,f(m,m)).
\end{equation*}

\noindent After replacing the bid with $f(b,m)$,

\begin{equation*}
    \arg\max_{b} \mathbb{E}_p \big[ (v - f(p,m)) \mathbbm{1}_{f(p,m) < f(b,m) \leq f(m,m)}\big]=z
\end{equation*}

\noindent where $f(z,m)=\min(v,f(m,m))$. If $\min(v,f(m,m))=v$, then $z=f^{-1}(v,m)$. On the other hand, if $\min(v,f(m,m))=f(m,m)$, then $z=m$. Since $f$ is strictly increasing, we have $z=\min(f^{-1}(v,m),m)$. Therefore, according to (\ref{EQProofoptimal bid}) the optimal bid is
\begin{equation*}
    b^*(m,v) = \min(f^{-1}(v,m),m)
\end{equation*}
\end{IEEEproof}

The specified optimal bid in Theorem~\ref{theorem1} depends on the value function (\ref{EQvaluefunc}) of the MDP. Therefore, in order to calculate the optimal bid, the SU needs to compute $U(m)$. Let $U^t$ be the value function at time $t$, we can find $U^t$ for $t=1, 2, \cdots $ iteratively as follows:

\begin{equation*}
    U^{t+1}(m)= \delta U^{t}(m) + \:\mathbb{E}\big[ v - (p + \delta (U^{t}(m) - U^{t}(m-p+a))) \big]^{+},
\end{equation*}

\noindent with the initial value of $U^0(m)=0$ for $\forall m$. It is worth noting that the above equation is another form of the value function defined in (\ref{EQvaluefunc}). If the SU loses the auction, the expectation term is zero in the above equation and the SU gets $\delta U^{t}(m)$. When the expectation term is positive and the SU wins the auction, $\delta U^{t}(m)$ terms cancel out and the SU gets $v - p + \delta U^{t}(m-p+a)$.

\section{Learning-based Optimal Bidding Strategy}\label{SecLearning}
In this section, we find an optimal bidding strategy of an SU without the i.i.d bids requirement. For this purpose, we formulate the bidding problem as a Markov game (also called a stochastic game)\footnote{The theory of MDP focuses on a single-user stationary environment. Game theory, on the other hand, studies the interaction of multiple users. Markov games extend game theory to MDP-like environments. In other words, Markov games generalize MDP to environments with multiple interacting users.} \cite{Littman01}.

An n-user stationary Markov game can be described by a tuple \mbox{$<S,~B_1,~\cdots,~B_n,~r_1,~\cdots,~r_n,~q>$} where $S$ is the state space, $B_i$ is the set of actions, $r_i$ is the reward function  for user $i$, $i=1,\cdots ,n$ and $q$ determines the state transition probabilities. Given state $s\in S$, each user independently chooses an action $b_i \in B_i$, and receives a reward $r_i$. Then, the state transitions to the next state based on transition probability function $q$ which follows the Markov property.

It is worth noting that in a Markov game, states are defined globally and for the environment. That is, all users make their decisions based on a common environment state, and the system state evolves as a result of joint actions. In accordance with Section~\ref{SecFormulation}, we consider a local state space $S_i$ for each SU $i$. We define the global state space as $S=S_1\times\cdots \times S_n$, and we let $S_{-i}=\times_{j\neq i} S_j$ be the joint state of all SUs other than $i$. The global state of the system at time slot $t$ is defined as $s^t=(s_i^t,s_{-i}^t)$.

Also, in such a Markov game, each SU reward depends on the global state and the joint action of all SUs. However, due to the distributed nature of wireless networks/markets, exact information about other SUs is not available. Therefore, an SU needs to learn about its opponents through observations made from participating in the auction.

It should be noted that, in contrast with \cite{Han11} that assumes SUs can stay out of the auction and monitor the results, we assume that an SU can make observations only through participating in the auction. Also, since the auction is sealed-bid, SUs cannot observe each other's bids, and no information is exchanged among SUs. Thus, we define the observation of an SU as its previous states, bids, and auction outcomes for that SU, in addition to the SU's current state. Formally, we define the observation of SU $i$ at time $t$ as $(s_i^{t''}, b_i^{t'}, o_i^{t'})$ for $t'=0, \cdots, t-1$ and $t''=0, \cdots, t$.

We utilize \emph{model-free reinforcement learning} approaches in which an SU learns its optimal bidding strategy without knowing the state transition probabilities. Q-learning \cite{Watkins1989learning,Watkins1992q} is a well-known example of model-free reinforcement learning algorithms. The main idea of Q-learning is to define a Q-function that represents the quality of a state-action pair. Then, for a given state, the optimal strategy would be to choose an action that gives the highest value for Q-function.

\subsection{State Space Classification}\label{subsecStateClassification}
In a Markov game, Q-functions are defined over the global state and joint actions of all SUs. However, as mentioned earlier SUs cannot observe states and actions of each other. Thus, SU $i$ needs to approximate the state of others $S_{-i}$. Since the winning threshold fully represents the state and behavior of other SUs, it suffices for an SU to keep an estimate of the winning threshold. Therefore, winning threshold can be used as the representative state of competing SUs. In order to reduce the time and space complexity of learning, we use a similar state classification as in \cite{Fu09} to classify the representative state space. Let $\mathcal{T}$ be the maximum value for the winning threshold. SU $i$ uniformly decomposes the range $[0, \mathcal{T}]$ into $N_i$ intervals as $[\mathcal{T}_0, \mathcal{T}_1), [\mathcal{T}_1, \mathcal{T}_2), \cdots, [\mathcal{T}_{N_i-1}, \mathcal{T}_{N_i}]$, where $\mathcal{T}_0 \leq \mathcal{T}_1 \leq \cdots \leq \mathcal{T}_{N_i} = \mathcal{T}.$

Depending upon the outcome of the auction, SU $i$ gets to know different information about its competitors. Let $\tilde{s}^t_{-i}$ be the approximated state of other SUs at time $t$, we have the following two cases:
\begin{enumerate}
  \item If SU $i$ wins the auction at time $t$, the winning threshold can be observed. Therefore, the representative state of other SUs is determined as
  \begin{equation*}
    \tilde{s}^t_{-i}=n, \;\;\; \text{if} \;\;\; p_t \in [\mathcal{T}_{n-1}, \mathcal{T}_{n})
  \end{equation*}

  \item When SU $i$ loses the current round of auction, the only information available to the SU is that its bid was lower than the winning threshold. Thus, the representative state of other SUs can be chosen as
  \begin{equation*}
    \tilde{s}^t_{-i}=n, \;\;\; \text{if} \;\;\; b_i^{t} \in [\mathcal{T}_{n-1}, \mathcal{T}_{n})
  \end{equation*}
\end{enumerate}

It is worth noting that the choice of $N_i$ leaves a tradeoff between complexity and performance for SU $i$. Higher values of $N_i$ results in more accurate approximation of $S_{-i}$, but at the cost of increased complexity.

\subsection{Transition Probability Estimation} \label{subsecTranApprox}
SU $i$ also needs to estimate the transition probabilities for representative state of other SUs. For this purpose, SU $i$ maintains an $N_i\times N_i$ matrix $Y$. Each element $y_{n,m}$ of the matrix indicates the number of transitions from $\tilde{s}^t_{-i}=n$ to $\tilde{s}^{t+1}_{-i}=m$. SU $i$ can update the matrix $Y$ through its observations and state space approximation described in previous subsection. Then, we can approximate the transition probabilities as follows:

\begin{equation*}
    q_{-i}(\tilde{s}^{t+1}_{-i}=m | \tilde{s}^t_{-i}=n)=\frac{y_{n,m}}{\sum_{m} y_{n,m}}
\end{equation*}

\subsection{The Learning Algorithm}
In this section, we present a learning-based bidding algorithm for an SU which depends only on the local observations of the SU. The learning algorithm is similar to the well-known Q-learning \cite{Watkins1992q} method, except that we include budget constraints of SUs, and we use state classification and transition probability approximation of other SUs, since the information about other SUs are limited in the network.

We define the Q-function of SU $i$ at time $t$ as follows. The quality of action $b_i$, when state of SU $i$ is $s_i$ and the representative state of others is $\tilde{s}_{-i}$, equals

\begin{equation}\label{Qfunction}
    Q_{i}^{t}(s_i, \tilde{s}_{-i}, b_i)=
    \begin{cases}
        (1-\alpha_{i}^{t}) Q_{i}^{t-1}(s_i, \tilde{s}_{-i}, b_i)+\alpha_{i}^{t}(r_i^{t}
         +\delta V_{i}^{t}(s_i, \tilde{s}_{-i})), \\
        \qquad\qquad\quad\qquad\quad\text{if}\; s_i^t=s_i, \tilde{s}^t_{-i}=\tilde{s}_{-i}, b_i^{t}=b_i\\ \\
    Q_{i}^{t-1}(s_i, \tilde{s}_{-i}, b_i) \qquad \qquad\qquad \text{Otherwise}
    \end{cases}
\end{equation}

\noindent where $0\leq \alpha_{i}^{t}<1$ is the SU's learning rate, $r_i^{t}$ is the immediate reward as defined in (\ref{equImReward}). The function $V_{i}^{t}(s_i, \tilde{s}_{-i})$ represents the value of the joint state $(s_i, \tilde{s}_{-i})$, which is the expected discounted utility starting from that state.

\begin{multline}\label{StateValueQ}
    V_{i}^{t}(s_i^{t}, \tilde{s}^t_{-i})= \displaystyle\sum_{s_i^{t+1}, \tilde{s}^{t+1}_{-i}}
    \bigg[ q_i(s_i^{t+1}| s_i^{t}, o_i^t) q_{-i}(\tilde{s}^{t+1}_{-i} | \tilde{s}^t_{-i})
     \max_{b_i\leq m_i^{t+1}} \big\{Q_{i}^{t-1}(s_i^{t+1}, \tilde{s}^{t+1}_{-i}, b_i) \big\}\bigg]
\end{multline}

In other words, the quality of a state-action pair (\ref{Qfunction}) is the immediate utility plus the discounted expected value of future states, and the value of a joint state (\ref{StateValueQ}) is the quality of the best action for that state. The results in \cite{Watkins1992q} show that the estimated values for $Q$ and $V$ converge to their true values if learning rates satisfy certain conditions. Therefore, if an SU learns the Q values, it can specify its optimal strategy, which is choosing the bid (action) with the highest Q value subject to its budget constraints. Thus, SU $i$ chooses its bid at time $t$ according to the following strategy:

\begin{equation}\label{OptimalbidQ}
    \pi_i^*(s_i^{t}, \tilde{s}^{t-1}_{-i})=\arg\max_{b_i\leq m_i^{t}} \Big\{\displaystyle\sum_{\tilde{s}^{t}_{-i}} q_{-i}(\tilde{s}^{t}_{-i} | \tilde{s}^{t-1}_{-i}) Q_{i}^{t-1}(s_i^{t}, \tilde{s}^{t}_{-i}, b_i) \Big\}
\end{equation}

\noindent The SU chooses a bid that maximizes its expected Q value, where the expectation is taken over the possible representative state of other SUs for the current time step. This is because SU $i$ can learn about other SUs' state only after bidding and observing the auction results. Given the information from previous time step and with the aid of transition probability approximation (Section~\ref{subsecTranApprox}), the SU can find the expected current state of other SUs.

The results in \cite{Singh2000} indicate that the greedy strategy that always chooses an action which maximizes the Q values may not provide enough exploration for the user to guarantee optimal performance. A very common approach is to add some randomness to the policy. We use $\epsilon$-greedy with decaying exploration in which, the SU chooses a random exploratory bid at the joint state $s$ with probability $\epsilon(s)=c/n(s)$, where $0<c<1$ and $n(s)$ is the number of times the joint state $s$ has been observed so far. The SU chooses the greedy Q-maximizing bid (i.e. (\ref{OptimalbidQ})) with probability of $1-\epsilon(s)$. In this approach the probability of exploration decays over time as the SU learns more.

\begin{algorithm}[!t]
\caption{Learning-based bidding for SU $i$}\label{AlgLearning}
\begin{algorithmic}[1]
\State Initialize the $Q_i$ values to zero for all possible states and bids
\State Initialize $n(s)$ values to zero for all possible joint states $s$
\For {Each time step $t$}
\State Observe the current state $s_i^{t}$
\State With probability $\epsilon(s_i^{t}, \tilde{s}^{t-1}_{-i})=c/n(s_i^{t}, \tilde{s}^{t-1}_{-i})$ choose a random bid, and with probability of $1-\epsilon(s_i^{t}, \tilde{s}^{t-1}_{-i})$ use the greedy strategy in (\ref{OptimalbidQ}) to place a bid
\State $n(s_i^{t}, \tilde{s}^{t-1}_{-i}) ++$
\State Observe the auction outcome $o_i^{t}$ and receive the immediate reward $r_i^{t}$
\State Estimate the state of other SUs $\tilde{s}^{t}_{-i}$ and update the corresponding transition probabilities as described in Sections~\ref{subsecTranApprox} and \ref{subsecStateClassification}
\State Compute the value of state $s_i^{t}, \tilde{s}^{t}_{-i}$ using (\ref{StateValueQ})
\State Update the Q value $Q_{i}^{t}(s_i^{t}, \tilde{s}^{t}_{-i}, b_i^{t})$ according to (\ref{Qfunction})
\EndFor
\end{algorithmic}
\end{algorithm}

The learning-based bidding algorithm for SU $i$ is summarized in Algorithm~\ref{AlgLearning}. The time complexity of the algorithm is dominated by learning state values (\ref{StateValueQ}) which can be done in $O(|S_i|\times N_i\times|B_i|)$, where $|S_i|$ is the state space size for SU $i$, $N_i$ is the number of classes for other SUs' states, and $|B_i|$ is the bid space for SU $i$. In terms of space complexity, the SU needs to keep a table of size $|S_i|\times N_i\times|B_i|$ for Q values.

\section{Numerical Results}\label{simulation}
In this section, we evaluate the performance of our proposed bidding algorithm. We compare our learning-based bidding algorithm (Algorithm~\ref{AlgLearning}) versus truthful bidding which is known to be the optimal bidding strategy without budget limits. When truthful bidding is used with budget constraints, an SU bids its true valuation when the budget allows, and bids zero if the remaining budget is lower than the true valuation. Since bidding algorithms intend to maximize utility of an SU, our performance metric of interest is the utility that an SU obtains over time.

\begin{figure}[!t]
\centering
\includegraphics[width=8.5cm,height=6.7cm]{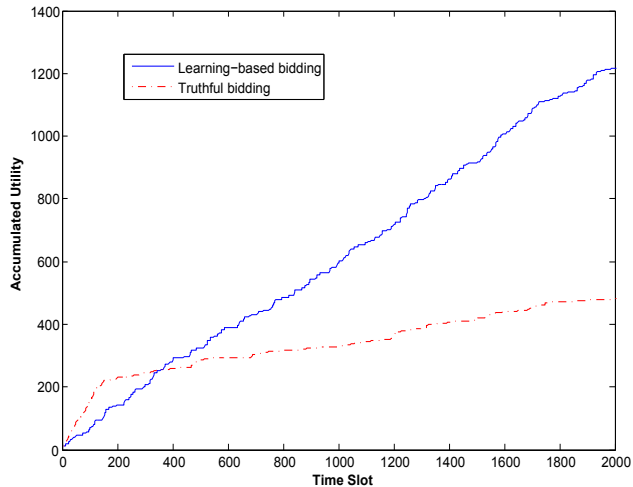}
\caption{The proposed learning-based bidding algorithm outperforms truthful bidding after the first 300 rounds.}
\label{fig:utility2000}
\end{figure}

The parameters in our numerical evaluations are set as follows. The SU starts with initial budget of 100, its valuation at each time slot is drawn randomly from discrete uniform distribution with maximum of 10, and SU's budget evolves according to (\ref{equStateTransition}). The discount factor $\delta$ is set to 0.8, the fixed income of SU for getting channel access, $a$, is 2, the learning rate $\alpha$ is constant over time and equals 0.5. We set the number of classes (intervals) for representing other SUs to $N=5$, and we choose 0.2 for the constant $c$ in Algorithm~\ref{AlgLearning}. The auction is repeated for 2000 rounds.

Fig.~\ref{fig:utility2000} shows the accumulated utility of an SU using our learning-based bidding versus truthful bidding. As can be seen, our proposed algorithm outperforms truthful bidding after the first 300 rounds. This is due to the fact that truthful bidding does not take into account budget planning. Therefore, the SU bids aggressively at first, which significantly reduces its remaining budget to the extent that the SU does not have enough competitive ability for the remaining auction rounds. On the other hand, our learning-based bidding method considers the effect of bids on the future and plans the budget wisely, which results in a better performance in the long run.

\begin{figure}[!t]
\centering
\includegraphics[width=8.5cm,height=6.7cm]{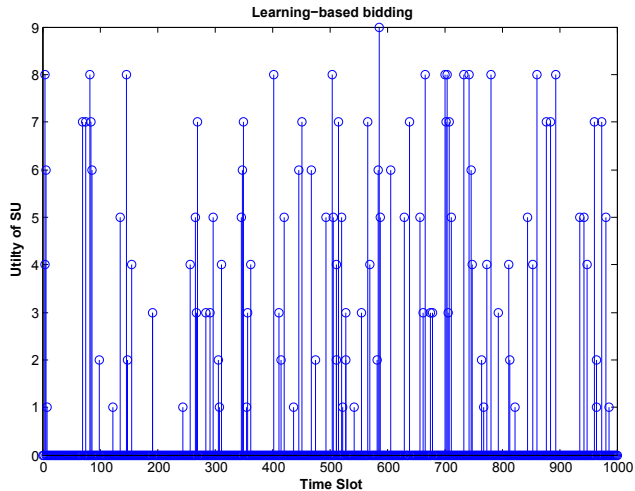}
\caption{The proposed learning-based bidding algorithm performs well after it learns about the competition in the beginning rounds of auction.}
\label{fig:utility}
\end{figure}

\begin{figure}[!t]
\centering
\includegraphics[width=8.5cm,height=6.7cm]{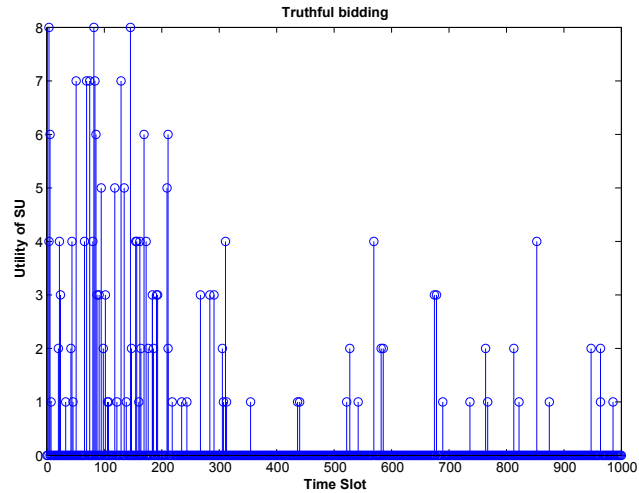}
\caption{Truthful bidding performs well for the first 200 rounds, but its performance degrades afterwards.}
\label{fig:utilityTruth}
\end{figure}

The utility of an SU using our learning-based bidding algorithm at each round of auction is shown in Fig.~\ref{fig:utility}. It can be seen that our proposed learning-based algorithm performs well after it learns about the competition in the beginning rounds of auction.
On the other hand, as Fig.~\ref{fig:utilityTruth} shows, the performance of the truthful bidding algorithm is only desirable for the first 200 rounds of auction. Although aggressive bidding in the truthful bidding algorithm brings large utilities at first, it leads to budget shortage very soon which consequently results in poor performance over time.

\begin{figure}[!t]
\centering
\includegraphics[width=8.5cm,height=6.7cm]{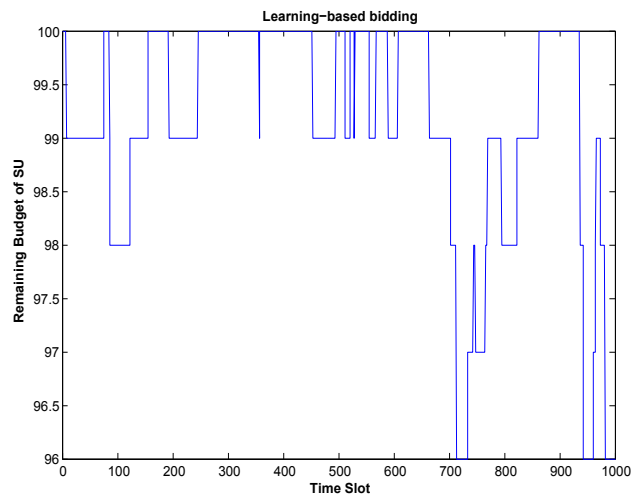}
\caption{Our learning-based bidding algorithm plans the budget wisely and maintains a good remaining budget over time.}
\label{fig:budget}
\end{figure}

\begin{figure}[!t]
\centering
\includegraphics[width=8.5cm,height=6.7cm]{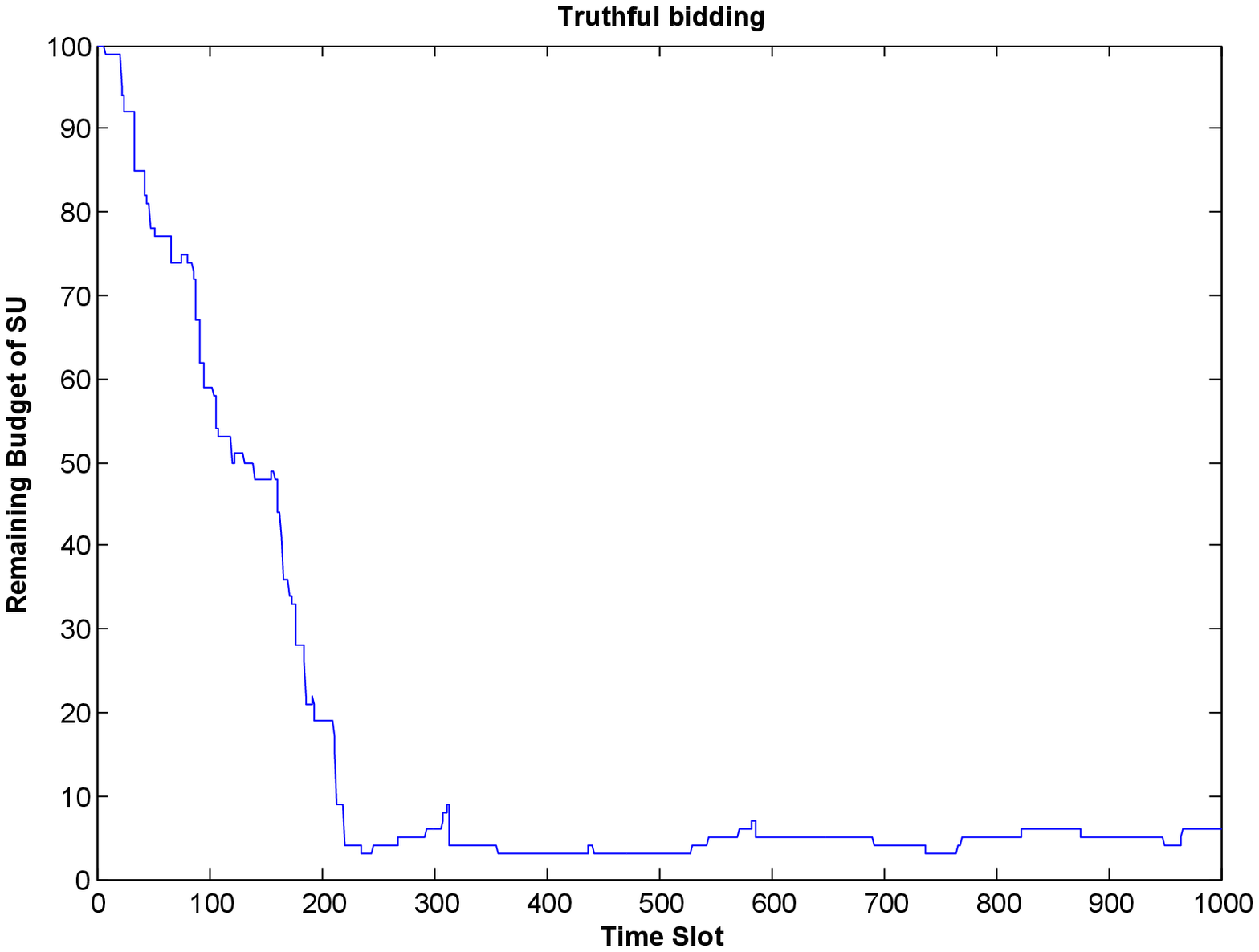}
\caption{Truthful bidding depletes its initial budget quickly, due to the lack of budget planning.}
\label{fig:budgetTruth}
\end{figure}

Fig.~\ref{fig:budget} and Fig.~\ref{fig:budgetTruth} show the evolution of an SU's budget over time using our learning-based bidding algorithm and truthful bidding, respectively. Fig.~\ref{fig:budget} illustrates that our learning-based bidding algorithm plans the budget wisely and maintains a good remaining budget over time. In contrast, the truthful bidding policy depletes the initial budget quickly which is due to its aggressive bidding style and lack of budget planning.

\section{Conclusion}\label{conclusion}
In this paper, we studied the bidding problem of a budget constrained SU in repeated secondary spectrum auctions. We presented an MDP formulation of the problem and characterized the optimal bidding strategy of an SU, assuming that opponents' bids are i.i.d. Then, we generalized the formulation to a Markov game that allows an SU to make inferences about its opponents based on local observations. Using model-free reinforcement learning approaches, we proposed a fully distributed learning-based bidding algorithm which relies only on local information. Through numerical evaluations, we showed that our learning-based bidding method outperforms truthful bidding, in terms of utility.

\section*{Acknowledgment}
This material is based upon work partially supported by the National Science Foundation under grant numbers 1422153 and 1456887.

\bibliography{BudgetingGame}

\begin{thebibliography}{10}
\providecommand{\url}[1]{#1}
\csname url@samestyle\endcsname
\providecommand{\newblock}{\relax}
\providecommand{\bibinfo}[2]{#2}
\providecommand{\BIBentrySTDinterwordspacing}{\spaceskip=0pt\relax}
\providecommand{\BIBentryALTinterwordstretchfactor}{4}
\providecommand{\BIBentryALTinterwordspacing}{\spaceskip=\fontdimen2\font plus
\BIBentryALTinterwordstretchfactor\fontdimen3\font minus
  \fontdimen4\font\relax}
\providecommand{\BIBforeignlanguage}[2]{{%
\expandafter\ifx\csname l@#1\endcsname\relax
\typeout{** WARNING: IEEEtran.bst: No hyphenation pattern has been}%
\typeout{** loaded for the language `#1'. Using the pattern for}%
\typeout{** the default language instead.}%
\else
\language=\csname l@#1\endcsname
\fi
#2}}
\providecommand{\BIBdecl}{\relax}
\BIBdecl

\bibitem{FCCRep}
{FCC Spectrum Policy Task Force}, \emph{Report of the spectrum efficiency
  working group}, Available: http://www.fcc.gov/sptf/reports.html, Nov. 2002.

\bibitem{Khaledi13}
M.~Khaledi and A.~Abouzeid, ``Auction-based spectrum sharing in cognitive radio
  networks with heterogeneous channels,'' in \emph{Information Theory and
  Applications Workshop (ITA), 2013}, 2013, pp. 1--8.

\bibitem{Hoefer15}
M.~Hoefer and T.~Kesselheim, ``Secondary spectrum auctions for symmetric and
  submodular bidders,'' \emph{ACM Trans. Econ. Comput.}, vol.~3, no.~2, pp.
  9:1--9:25, Apr. 2015.

\bibitem{Han11}
Z.~Han, R.~Zheng, and H.~Poor, ``Repeated auctions with bayesian nonparametric
  learning for spectrum access in cognitive radio networks,'' \emph{Wireless
  Communications, IEEE Transactions on}, vol.~10, no.~3, pp. 890--900, March
  2011.

\bibitem{Akkarajitsakul11}
K.~Akkarajitsakul, E.~Hossain, and D.~Niyato, ``Distributed resource allocation
  in wireless networks under uncertainty and application of bayesian game,''
  \emph{Communications Magazine, IEEE}, vol.~49, no.~8, pp. 120--127, August
  2011.

\bibitem{Ji08}
Z.~Ji and K.~J.~R. Liu, ``Multi-stage pricing game for collusion-resistant
  dynamic spectrum allocation,'' \emph{IEEE Journal on Selected Areas in
  Communications}, vol.~26, no.~1, pp. 182--191, Jan 2008.

\bibitem{Bae08}
J.~Bae, E.~Beigman, R.~Berry, M.~Honig, and R.~Vohra, ``Sequential bandwidth
  and power auctions for distributed spectrum sharing,'' \emph{Selected Areas
  in Communications, IEEE Journal on}, vol.~26, no.~7, pp. 1193--1203,
  September 2008.

\bibitem{Duan10}
L.~Duan, J.~Huang, and B.~Shou, ``Competition with dynamic spectrum leasing,''
  in \emph{New Frontiers in Dynamic Spectrum, 2010 IEEE Symposium on}, April
  2010, pp. 1--11.

\bibitem{Niyato08}
D.~Niyato and E.~Hossain, ``Competitive pricing for spectrum sharing in
  cognitive radio networks: Dynamic game, inefficiency of nash equilibrium, and
  collusion,'' \emph{Selected Areas in Communications, IEEE Journal on},
  vol.~26, no.~1, pp. 192--202, Jan 2008.

\bibitem{Niyato09}
D.~Niyato, E.~Hossain, and Z.~Han, ``Dynamics of multiple-seller and
  multiple-buyer spectrum trading in cognitive radio networks: A game-theoretic
  modeling approach,'' \emph{Mobile Computing, IEEE Transactions on}, vol.~8,
  no.~8, pp. 1009--1022, 2009.

\bibitem{Min12}
A.~Min, X.~Zhang, J.~Choi, and K.~Shin, ``Exploiting spectrum heterogeneity in
  dynamic spectrum market,'' \emph{Mobile Computing, IEEE Transactions on},
  vol.~11, no.~12, pp. 2020--2032, 2012.

\bibitem{Cramton97}
\BIBentryALTinterwordspacing
P.~Cramton, ``The fcc spectrum auctions: An early assessment,'' \emph{Journal
  of Economics and Management Strategy}, vol.~6, no.~3, pp. 431--495, 1997.
  [Online]. Available: \url{http://dx.doi.org/10.1111/j.1430-9134.1997.00431.x}
\BIBentrySTDinterwordspacing

\bibitem{FCCRep35GHz}
{FCC}, \emph{{Amendment of the Commission's Rules with Regard to Commercial
  Operations in the 3550-3650 MHz Band}}, {Available:
  https://apps.fcc.gov/edocs\_public/attachmatch/FCC-15-47A1.pdf}, Apr. 2015.

\bibitem{Fu09}
F.~Fu and M.~van~der Schaar, ``Learning to compete for resources in wireless
  stochastic games,'' \emph{Vehicular Technology, IEEE Transactions on},
  vol.~58, no.~4, pp. 1904--1919, May 2009.

\bibitem{Borgs07}
C.~Borgs, J.~Chayes, N.~Immorlica, K.~Jain, O.~Etesami, and M.~Mahdian,
  ``Dynamics of bid optimization in online advertisement auctions,'' in
  \emph{Proceedings of the 16th international conference on World Wide
  Web}.\hskip 1em plus 0.5em minus 0.4em\relax ACM, 2007, pp. 531--540.

\bibitem{Feldman07}
J.~Feldman, S.~Muthukrishnan, M.~Pal, and C.~Stein, ``Budget optimization in
  search-based advertising auctions,'' in \emph{Proceedings of the 8th ACM
  conference on Electronic commerce}.\hskip 1em plus 0.5em minus 0.4em\relax
  ACM, 2007, pp. 40--49.

\bibitem{Amin12}
K.~Amin, M.~Kearns, P.~Key, and A.~Schwaighofer, ``Budget optimization for
  sponsored search: Censored learning in mdps,'' in \emph{Proceedings of the
  Twenty-Eighth Conference on Uncertainty in Artificial Intelligence, Catalina
  Island, CA, USA, August 14-18, 2012}, 2012, pp. 54--63.

\bibitem{Gummadi13}
R.~Gummadi, P.~B. Key, and A.~Proutiere, ``Optimal bidding strategies in
  dynamic auctions with budget constraints,'' in \emph{Communication, Control,
  and Computing (Allerton), 2011 49th Annual Allerton Conference on}, Sept
  2011, pp. 588--588.

\bibitem{Khaledi15}
M.~Khaledi and A.~Abouzeid, ``Dynamic spectrum sharing auction with
  time-evolving channel qualities,'' \emph{Wireless Communications, IEEE
  Transactions on}, vol.~14, no.~11, pp. 5900--5912, Nov 2015.

\bibitem{AGT}
N.~Nisan, T.~Roughgarden, E.~Tardos, and V.~V. Vazirani, \emph{Algorithmic Game
  Theory}.\hskip 1em plus 0.5em minus 0.4em\relax New York, NY, USA: Cambridge
  University Press, 2007.

\bibitem{Littman01}
M.~L. Littman, ``Value-function reinforcement learning in markov games,''
  \emph{Cognitive Systems Research}, vol.~2, no.~1, pp. 55 -- 66, 2001.

\bibitem{Watkins1989learning}
C.~J. C.~H. Watkins, ``Learning from delayed rewards,'' Ph.D. dissertation,
  University of Cambridge England, 1989.

\bibitem{Watkins1992q}
C.~J. Watkins and P.~Dayan, ``Q-learning,'' \emph{Machine learning}, vol.~8,
  no. 3-4, pp. 279--292, 1992.

\bibitem{Singh2000}
S.~Singh, T.~Jaakkola, M.~L. Littman, and C.~Szepesv\'{a}ri, ``Convergence
  results for single-step on-policyreinforcement-learning algorithms,''
  \emph{Mach. Learn.}, vol.~38, no.~3, pp. 287--308, Mar. 2000.

\end{thebibliography}
\bibliographystyle{IEEEtran}

\end{document}